\documentclass[twocolumn,showpacs,floatfix,superscriptaddress,prc]{revtex4}
\usepackage{graphicx}
\usepackage{dcolumn}
\usepackage{bm}

\usepackage[normalem]{ulem}  
\usepackage[dvips]{color} 

\begin{document}

\title{Equation of state of the hot dense matter in a multi-phase transport model}
\author{Bin Zhang}
\affiliation{Department of Chemistry and Physics, Arkansas State University, State
University, AR 72467-0419, USA}
\author{Lie-Wen Chen}
\affiliation{Institute of Theoretical Physics, Shanghai Jiao Tong University, Shanghai
200240, China}
\author{Che Ming Ko}
\affiliation{Cyclotron Institute and Physics Department, Texas A\&M University, College
State, TX 77843-3366, USA}
\date{May 27, 2007}

\begin{abstract}
Within the framework of a multi-phase transport model, we study the equation
of state and pressure anisotropy of the hot dense matter produced in central
relativistic heavy ion collisions. Both are found to depend on the
hadronization scheme and scattering cross sections used in the model.
Furthermore, only partial thermalization is achieved in the produced matter
as a result of its fast expansion.
\end{abstract}

\maketitle

\section{Introduction}

Many experimental results have been obtained from the collisions of
heavy nuclei at the Relativistic Heavy Ion Collider (RHIC). To
understand these results, both macroscopic hydrodynamic models
\cite{Teaney:2000cw, Kolb:2001qz,Huovinen:2001cy,Hirano:2005xf} and
microscopic transport models
\cite{Zhang:1997ej,Molnar:2001ux,Bass:2002fh,Molnar:2004yh,Xu:2004mz}
have been used. Assuming that a thermalized quark-gluon plasma is
produced very early during the collisions, the ideal hydrodynamic
model without viscosity is able to reproduce measured transverse
momentum spectra of various hadrons and their large elliptic flows.
These experimental data can also be described by transport models
that include a partonic phase in which partons undergo scatterings
with cross sections much larger than those given by the perturbative
QCD. Although the transport model approaches the ideal hydrodynamic
model when the mean free path of a particle is much smaller than
both the inter-particle distances and the size of the colliding
system, it is not clear if these conditions are satisfied in
relativistic heavy ion collisions. Studies have thus been carried
out to understand how results from microscopic transport models
differ from the macroscopic hydrodynamic models, particularly for
the elliptic flow developed in relativistic heavy ion collisions
\cite{Molnar:2004yh,Gombeaud:2007ub}. Other questions which are also
of great interest are the equation of state (EOS) of and
the degree of thermalization in the hot dense
matter described by these models.

In the present paper, we study the EOS and pressure anisotropy of the hot
dense matter produced in central relativistic heavy ion collisions in A
Multi-Phase Transport (\textsc{AMPT}) model. The results indicate that both
the EOS and pressure anisotropy depend on the hadronization scheme and
scattering cross sections used in the model. Furthermore, only partial
thermalization is achieved in the produced matter as a result of its fast
expansion.

This paper is organized as follows. In Sec. \ref{ampt}, the AMPT model is
briefly reviewed. Results on the EOS in the AMPT model are shown in Sec. \ref%
{eos} while the pressure anisotropy is studied in Sec. \ref{pressure}.
Finally, a brief summary and discussions are given in Sec. \ref{summary}.

\section{The AMPT model}

\label{ampt}

The AMPT model is a hybrid model \cite%
{Zhang:1999bd,Lin:2000cx,Lin:2001yd,Lin:2004en}. It makes use of
different models for different stages of relativistic heavy ion
collisions. The initial conditions are taken from the Heavy Ion Jet
Interaction Generator (\textsc{HIJING}) \cite{Wang:1991ht}. In the
default AMPT model, the partonic matter consists of only mini-jet
partons from \textsc{HIJING}. It is different in the string melting
scenario in which the hadrons generated by \textsc{HIJING} are
dissociated according to their valence quark structures and
the resulting partonic matter is thus much denser.
The evolution of the partonic matter is simulated with Zhang's
Parton Cascade (\textsc{ZPC}) \cite{Zhang:1997ej}. At the moment,
only elastic parton scatterings are included in \textsc{ZPC} with
the cross sections regulated by a screening mass. After partons stop
scattering, they are converted into hadrons. In the default
\textsc{AMPT} model, partons are combined first with their parent
strings which then fragment according to the Lund model as
implemented in \textsc{PYTHIA} \cite{Sjostrand:1993yb}. In the
\textsc{AMPT} model with string melting, closest partons are
converted into hadrons via the coalescence model
\cite{Greco:2003xt,Greco:2003mm}. The resulting hadronic system
evolves according to A Relativistic Transport (\textsc{ART}) model
\cite{Li:1995pr,Li:2001xh}. The default \textsc{AMPT} model can give
a good description of particle distributions
\cite{Lin:2000cx,Lin:2001yd} while the \textsc{AMPT} model with
string melting is needed to describe the elliptic flow
\cite{Lin:2001zk} and HBT radii \cite{Lin:2002gc}. Other observables
have also been studied within the framework of the \textsc{AMPT}
model, and these include phi meson production \cite{Pal:2002aw},
higher-order flows \cite{Chen:2004dv}, pseudo-rapidity
\cite{Chen:2004vh} and system-size \cite{Chen:2005mr,Chen:2005zy}
dependence of flows, charmonium production
\cite{Zhang:2000nc,Zhang:2002ug} as well as charm
\cite{Zhang:2005ni} and strange \cite{Chen:2006vc} hadron flows.

\section{Equation of state in the AMPT model}

\label{eos}

The equation of state of a matter can be described by the relation
between its pressure and energy density. Since only particles
contribute to the pressure and energy density in the \textsc{AMPT}
model, the pressure and energy density can be extracted from the
energy-momentum tensor
\begin{equation}
T^{\mu \nu }(\bm{r})=\int \frac{d^{3}p}{(2\pi )^{3}}\frac{p^{\mu }p^{\nu }}{E%
}f(\bm{r},\bm{p}),
\end{equation}%
where $f(\bm{r},\bm{p})$ is the particle phase-space distribution
function. When the latter is sampled by point particles, the
energy-momentum tensor can be calculated by
averaging over particles and events in a volume $V$, i.e.,
\begin{equation}
T^{\mu \nu }=\frac{1}{V}\sum_{i}\frac{p_{i}^{\mu }p_{i}^{\nu
}}{E_{i}}.
\end{equation}%
In Cartesian coordinates where the $z$-axis points along the colliding
beam direction,
the pressure components are related to the energy-momentum tensor by $%
P_{x}=T^{11}$, $P_{y}=T^{22}$, $P_{z}=T^{33}$, and the energy
density is given by $\epsilon =T^{00}$. In central collisions,
because of the cylindrical symmetry around the beam axis,
$P_{x}=P_{y}$, the transverse pressure can be defined to be
$P_{T}=\frac{1}{2}(P_{x}+P_{y})$. The longitudinal pressure $P_{L}$
is the same as $P_{z}$. For a system in thermal equilibrium, its
pressure is isotropic, i.e., $P_{T}=P_{L}=P$. Otherwise, we define
the total pressure as $P=\frac{1}{3}(P_{x}+P_{y}+P_{z})$. The
equation of state of the system is then
described by $P/\epsilon $ as a function of $\epsilon $. In the
following, calculations will be carried out 
for central Au+Au collisions at $\sqrt{s_{NN}}=200$
GeV. In particular, we consider the central cell
of produced matter, specified by the space-time
rapidity $\eta_s$ and the transverse radial distance $r$. Only
active particles, i.e., particles that have been formed but not yet
frozen out, will be counted and contributions from free streaming
particles are not included.

\begin{figure}[!htb]
\includegraphics[scale=0.65]{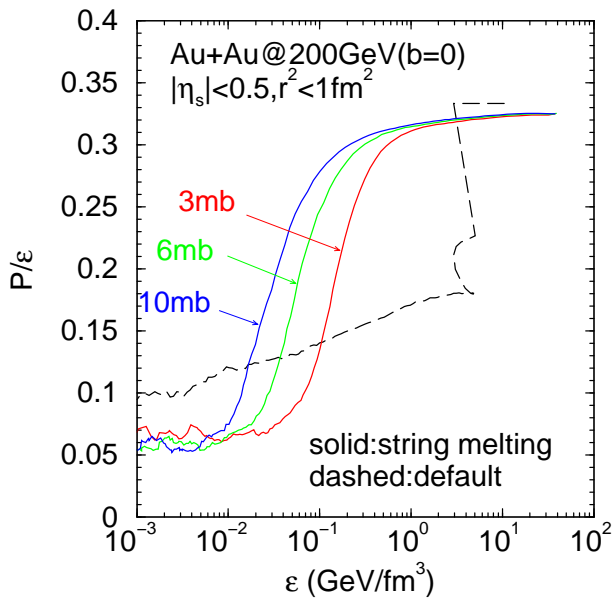}
\caption{(Color online) Equation of state in the
\textsc{AMPT} model.} \label{poe}
\end{figure}

Fig.~\ref{poe} shows the equation of state extracted from the
default \textsc{AMPT} model as well as the \textsc{AMPT} with string
melting for three different parton scattering cross sections. In the
initial partonic phase, the default \textsc{AMPT} gives a slightly
higher $P/\epsilon $ than the string melting \textsc{AMPT} because
the partonic phase in the default model mainly consists
of massless gluons 
while that in the string melting model is dominated by light
quarks with non-zero quark masses. Also, there are fewer partons in
the default model than in the string melting model. Furthermore,
hadronization happens earlier in the default model when $\epsilon $
is around a few GeV while in the string melting model hadronization
starts when $\epsilon $ is an order of magnitude smaller. The
partonic scattering cross section also affects the equation of state
and the transition between the partonic phase and the hadronic
phase. When the cross section in the string melting model increases
from $3$ mb, to $6$ mb, and then to $10$ mb, the equation of state
becomes harder and reaches into regions of lower
energy density. In the hadronic
stage, the default model has a harder equation of state compared to
that in the string melting model. The two models thus give very
different pressures for the same energy density in the hadronic
stage.

\begin{figure}[!htb]
\includegraphics[scale=0.65]{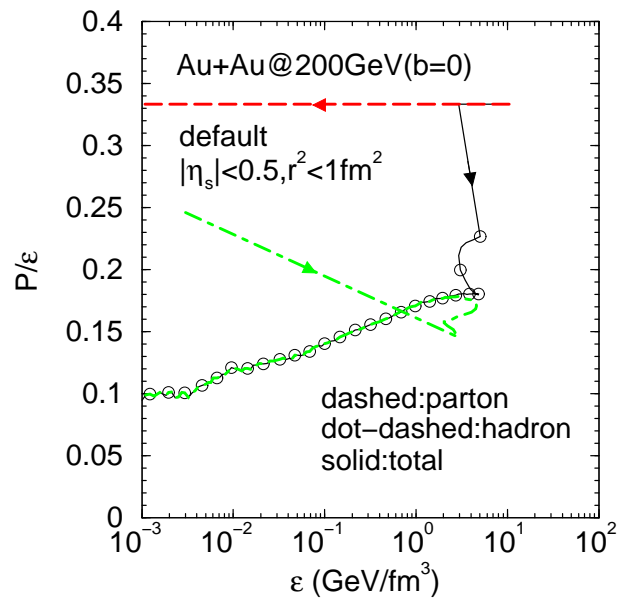}
\caption{(Color online) Equation of state in the partonic phase and
the hadronic phase as well as combined partonic and hadronic matters
from the default AMPT model. Arrows indicate directions of
increasing proper time. The circles on the solid
curve are data points taken at $\tau=1, 2, 3, \dots$ fm/c.
} \label{poe_df}
\end{figure}

The equation of state in the default model shows a sharp transition
as a result of combining the equation of state of the partonic phase
with that of the hadronic phase as shown in Fig.~\ref{poe_df}. In
the default model, hadron production is related to the parent
strings. The formation proper time is the average parton freeze-out
time plus $0.7$ fm/c. As a result, hadron production is
earlier for strings without final partonic interactions compared
with those that involve final partonic interactions. These early
hadrons lead to the increase in hadron energy density from about
$3\times 10^{-3}$ GeV/fm$^{3}$ at $0.8$ fm/c when the system is
still in the parton phase. Hadrons from strings with final state
interactions lead to a second increase from about $2$ GeV/fm$^{3}$
at $1.8$ fm/c as seen in Fig.~\ref{poe_df}. The sudden change in the
equation of state is due to the transformation of the string energy,
which is not included in the calculations of the energy density and
pressure, into hadrons during string fragmentation. This process
completes at about 3 fm/c when the hadronic stage takes over.

\begin{figure}[!htb]
\includegraphics[scale=0.65]{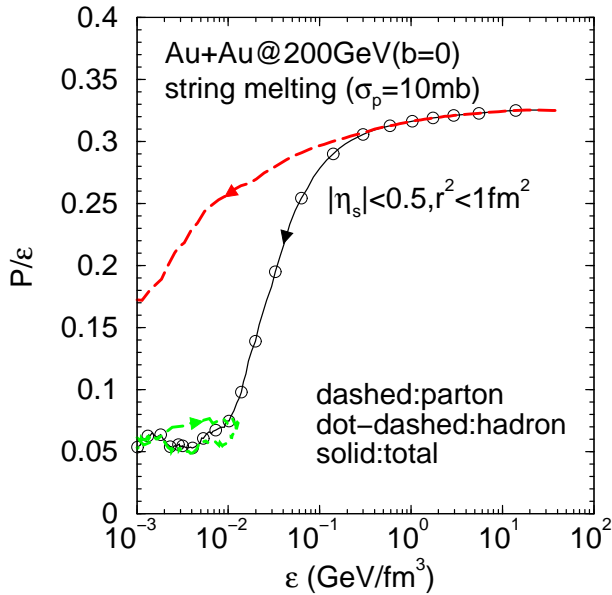}
\caption{(Color online) Same as Fig.~\protect\ref{poe_df} from the
AMPT model with string melting.
} \label{poe_sm}
\end{figure}

The equation of state of the partonic phase as well as that of the
hadronic phase in the string melting model with a parton scattering
cross section of $10$ mb are shown in Fig.~\ref{poe_sm}. Unlike the
default model in which gluons dominate the partonic stage, only
quark degrees of freedom are included in the string melting model.
As light quarks move out of the central cell, the relative abundance
of strange quarks in the central cell increases, leading to a
decrease of $P/\epsilon $ as $\epsilon $ decreases. The system stays
in the partonic stage until about $7$ fm/c when hadrons begin to
affect the equation of state. The hadron energy density at this time
is about $10^{-3}$ GeV/fm$^3$. As hadrons are formed from coalescing
partons, their contribution to the energy increases. Although the
energy density decreases as a result of the expansion of the system,
the pressure drops even faster during hadronization. The equation of
state of the whole system thus gradually changes from the partonic
equation of state to the hadronic equation of state. Hadrons begin
to play an important role when the hadron energy density reaches its
maximum value of about $10^{-2}$ GeV/fm$^3$ at about $10$ fm/c and
they dominate the energy density at about $13$ fm/c. It is seen from
Fig.~\ref{poe_df} and Fig.~\ref{poe_sm} that the equation of state
of the hadronic phase may have two different values for the pressure
at a fixed energy density. This feature reflects the fact that the
hadronic phase experiences the formation stage and expansion stage
in the collisions, and the two stages have different particle
compositions and/or momentum distributions and hence different
pressures.

\section{Pressure anisotropy in the AMPT model}

\label{pressure}

\begin{figure}[htbp]
\includegraphics[scale=0.65]{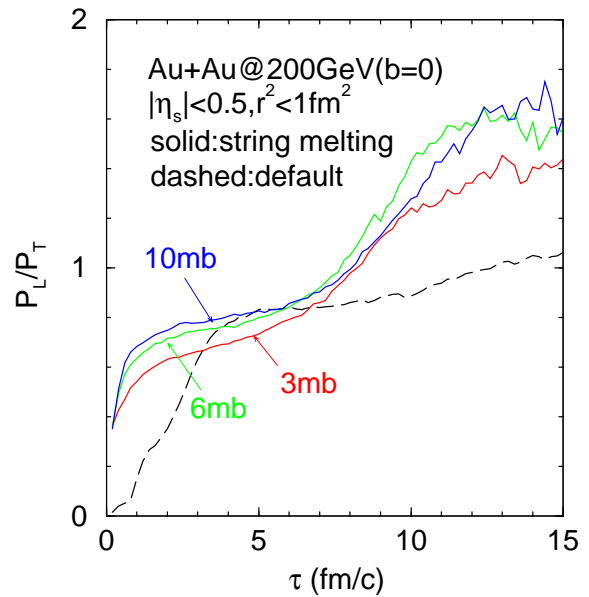}
\caption{(Color online) Proper time evolution of the pressure
anisotropy parameter in the \textsc{AMPT} model.}
\label{plopt}
\end{figure}

In the above calculations, the energy density and pressure are
obtained from the energy-momentum tensor, and they
are not necessarily for a matter in thermal equilibrium. In order to
check if the matter formed in relativistic heavy ion collisions is
thermalized, we show in Fig.~\ref{plopt} the proper time evolution
of the pressure anisotropy parameter $P_{L}/P_{T}$ defined by the
ratio of the longitudinal to the transverse pressures. In the
default model both the number of partons and the number of
collisions are small during the early stage, and the anisotropy
parameter starts to increase quickly only when hadrons come into
play at about $2$ fm/c. In contrast, in the string melting model
there is a quick increase in $P_{L}/P_{T}$ and by $1$ fm/c its value
is already over $0.5$. With increasing parton cross section, the
system becomes more equilibrated and $P_{L}/P_{T}$ becomes also
larger. It is interesting to note that $P_{L}/P_{T} $ crosses $1$ at
some time. However, it does not stay at $1$ for any significant
period of time. In the string melting model, this happens after
about $6$ fm/c where there is a change from the gradual saturation
of $P_{L}/P_{T}$ to a rapid increase, while in the default model,
the change happens at about $10$ fm/c. The proper time evolution of
the energy density also shows a change at about the same time when
the effect of transverse expansion sets in. After this time,
particles with large transverse momentum move out of the central
cell, leading to an increase in $P_{L}/P_{T}$. Therefore, to have
$P_{L}/P_{T}=1$ temporarily does not indicate full thermalization
during the evolution of the system. Furthermore, $P_{L}/P_{T}$ in
the central cell can even go beyond $1$. If only the initial stage
is considered, the anisotropy parameter can only reach a saturation
value of about $0.8$, implying that only partial thermalization is
achieved in the central cell. We note that the present results are
consistent with those obtained in Ref. \cite{Xu:2004mz} with large
two-body parton cross sections where the momentum anisotropy
has been shown for the initial stage
of central Au+Au collisions. It is also interesting
to note that while larger cross section makes initial pressure
anisotropy closer to $1$, it also leads to 
delays both in the transverse expansion and 
in the rise of resulting pressure anisotropy.
This is what underlies the
crossing of the 6 mb and 10 mb curves in Fig.~\ref{plopt}.

\begin{figure}[!htb]
\includegraphics[scale=0.65]{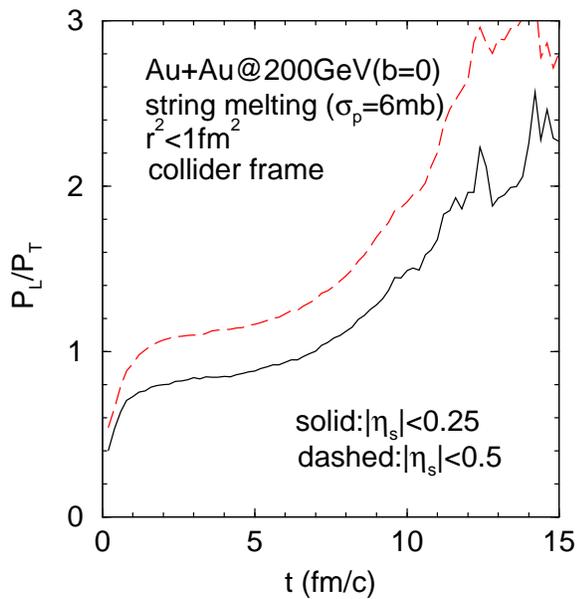}
\caption{(Color online) Time evolution of pressure anisotropy in the string
melting model for two different space-time rapidity bin widths in the
collider frame.}
\label{plopt_t}
\end{figure}

\begin{figure}[!htb]
\includegraphics[scale=0.65]{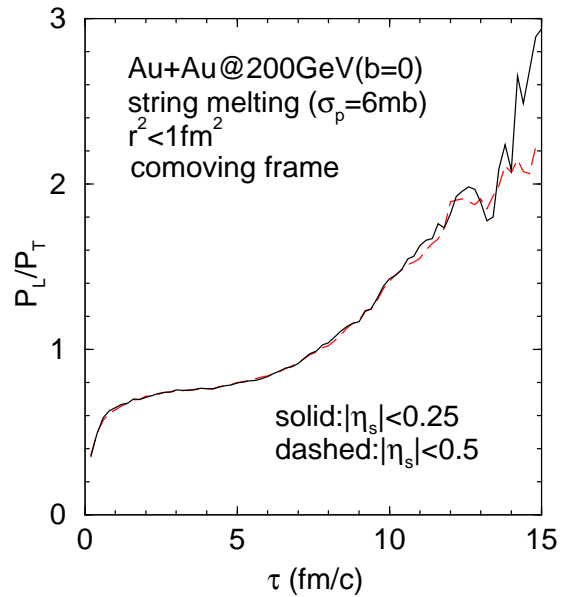}
\caption{(Color online) Same as Fig.~\protect\ref{plopt_t}
but for proper time evolution in the comoving frame.}
\label{plopt_tau}
\end{figure}

For an accurate evaluation of the pressure anisotropy, momentum in the local
rest frame needs to be used even for a space-time rapidity bin within $|\eta
_{s}|\leq 0.5$. Specifically, the local momentum $p_{z}^{\prime }$ can be
calculated from energy $E$ and momentum $p_{z}$ by%
\begin{equation}
p_{z}^{\prime }=\frac{t\,p_{z}-z\,E}{\tau }
\end{equation}%
with the proper time $\tau =\sqrt{t^{2}-z^{2}}$
\cite{Baym:1984np,Gavin:1990up}. The evolution of the anisotropy
parameter evaluated with the energy and momentum in the lab or
collider frame is shown in Fig.~\ref{plopt_t}. The results depend
strongly on the space-time rapidity bin width. However, if the local
energy and momentum are used, i.e., in the comoving frame, the
dependence on the space-time rapidity width disappears as shown in
Fig.~\ref{plopt_tau}. This indicates that the longitudinal flow is
boost invariant around the mid-space-time rapidity.

\section{Summary and discussions}

\label{summary}

We have studied the equation of state in the
AMPT model and found it
to depend on both the hadronization scheme and the parton scattering
cross sections. In the default model, the equation of state is much
softer than that in the
string melting model over a wide range of energy density
as a result of early hadronization. In the
latter case, hardness of the equation of state increases with
increasing parton cross section. In both scenarios, the produced hot
dense matter is not in full thermal equilibrium as the pressure
isotropy deviates appreciably from unity during most of its
evolution as a result of fast expansion. The non-equilibrium nature
of the equation of state in the \textsc{AMPT} model thus differs
from that commonly used in hydrodynamic models and also that
obtained from lattice Quantum-Chromodynamics simulations
\cite{Ejiri:2005uv}. There are recent attempts of improving the
current description in the \textsc{AMPT} model by imposing
hadronization at fixed time \cite{Meiling:2006dq}. This certainly
can make the equation of state at later time softer. Mean field can
also modify the equation of state \cite{Tan:2006uq}. Also, an
important aspect of the \textsc{AMPT} model is that the partonic
phase includes only elastic scatterings. In a recent parton cascade
model that includes parton number changing processes
\cite{Xu:2004mz}, the partonic matter is able to reach pressure
isotropy. Parton inelastic scatterings are therefore important in
bringing the system closer to ideal hydrodynamic solutions, which
require at least pressure isotropy to be reached and maintained
\cite{Heinz:2005zi}. Furthermore, plasma instabilities may help the
isotropization process as well \cite%
{Dumitru:2005gp,Arnold:2005vb,Rebhan:2005re,Romatschke:2006nk}. The
consequences of including these mechanisms in the AMPT model on both the
equation of state of produced hot dense matter and other physical
observables deserve further investigations.

\begin{acknowledgments}
We acknowledge S. Bass, T. Renk, and U. Heinz for helpful discussions. B.Z.
thanks the hospitality of the National Institute for Nuclear Theory during
the INT-03 program where part of the work was done. The calculations of the
work were performed using computer resources provided by the Parallel
Distributed Systems Facilities of the National Energy Research Scientific
Computing Center. The work was supported by the U.S. National Science
Foundation under grant No.'s PHY-0554930 and PHY-0457265, the Welch
Foundation under grant No. A-1358, the NNSF of China under Grant Nos.
10575071 and 10675082, MOE of China under project NCET-05-0392, Shanghai
Rising-Star Program under Grant No. 06QA14024, and the SRF for ROCS, SEM of
China.
\end{acknowledgments}

\end{document}